# Multicriteria Analysis Model in Sustainable Corn Farming Area Planning

Abdul Haris[1], Muhammad Munawir Syarif[1], Hamed Narolla[2], Rachmat Hidayat[1]

[1]Agrotechnology, Faculty of Agriculture, Universitas Muslim Indonesia, Indonesia
[2]Agricultural Economics and Rural Development, Faculty of Agriculture, Omdurman Islamic University, Sudan
Email: abdul.haris@umi.ac.id

Abstract

This study aims to develop a framework for multicriteria analysis to evaluate alternatives for sustainable corn agricultural area planning, considering the integration of ecological, economic, and social aspects as pillars of sustainability. The research method uses qualitative and quantitative approaches to integrate ecological, economic, and social aspects in the multicriteria analysis. The analysis involves land evaluation, subcriteria identification, and data integration using Multidimensional Scaling and Analytical Hierarchy Process methods to prioritize developing sustainable corn agricultural areas. Based on the results of the RAP-Corn analysis, it indicates that the ecological dimension depicts less sustainability. The AHP results yield weight distribution and highly relevant scores that describe tangible preferences. Priority directions are grouped as strategic steps toward achieving the goals of sustainable corn agricultural area planning.

## INTRODUCTION

Corn farming is one of the vital sectors in the Indonesian economy [1]. Bengo District is one of the areas in Bone Regency with the highest corn production in South Sulawesi [2]. Corn agricultural areas play an important role in providing food security and increasing farmers' income. However, developing planning for these areas cannot rely solely on one aspect but must consider various complex and sustainable factors [3]. The pillars of sustainable development are based on ecological, economic, and social aspects [4].

Multicriteria analysis assists in evaluating decision support systems for assessing alternative agricultural area planning [5]. Spatial multicriteria analysis integrates geographic data and value assessments to enable balanced decision-making across various parameters, encompassing economic, social, technical, and environmental aspects [6, 7]. This approach offers advantages in providing comprehensive assessments of relevant aspects. Challenges of this approach include complexity, difficulty in integration, and the subjective element of criterion importance levels [8, 9], thus necessitating multicriteria techniques for multivariate systems [10].

Multidimensional Scaling (MDS) is a multivariate analysis technique that can help in visualizing the relationship between alternatives and criteria in multidimensional spaces [11, 12] which allows us to visually understand issues on ecological, economic, and social aspects of corn farming planning. The Analytic Hierarchy Process (AHP) is a decision-making method to solve complex problems by separating several criteria and subcriteria to measure quantitative preferences of criteria [13,14]. The combination of MDS and AHP in multiple criteria decision analysis will provide a holistic and structured approach to evaluating more informed and data-based planning alternatives.

The study provides a concrete overview and valuable insight into the application of a multi-criterion model in evaluating sustainable corn area planning alternatives through a combination of MDS and AHP. The main objective of the study is to develop a framework for multi-criterion analysis that can be used to evaluate alternatives to corn agricultural area planning, considering the integration of ecological, economic, and social aspects as a pillar of sustainability.

## RESEARCH METHODS

This research was conducted in Bengo District, Bone Regency, South Sulawesi Province from March to November 2023. Primary data was obtained through field surveys and a series of interviews with stakeholders/experts. This research uses qualitative and quantitative approaches in integrating various aspects of land, ecological, economics and social diversity in a multi-criteria analysis. This research design includes stages starting from land evaluation, identification of sub-criteria to data analysis and formulation of recommendations for planning a sustainable corn farming area.

**Land Suitability Analysis of Corn Plants**

The purpose of this analysis is to determine the suitability of land by matching land characteristics with the criteria of corn growing conditions. The suitability classification of land is based on the criterion of the Food and Agriculture Organization (FAO) [15].

**Identification of Sub-criteria and Analysis of Integration of Ecological, Economic and Social Aspects**

The relevant subcriteria will be collected and determined through the literature review related to sustainable corn farming planning, which will cover sustainability aspects of the ecological, social and economic dimensions.

The results of the land suitability analysis of corn crops are then collected subcriteria data from the ecological, social and economic aspects of the sustainability of corn farming. The data that has been collected will be further analyzed using statistical methods and multivariate analysis techniques through MDS. The analysis is carried out with the ordination technique called *RAP-Corn* (*Rapid Appraisal for Corn*) modified from *RAPFISH* [16]. This analysis is an integration analysis to produce a holistic assessment of alternative planning of sustainable corn farming areas. The results of the MDS analysis will give an index and, sustainability status value on ecological, economic and social aspects.

**Table 1.** Index Categories and Sustainability Status [16]

| Index Value | Category |
|---|---|
| 0.00 – 25.00 | Not Sustainable |
| 25.01 – 50.00 | Less Sustainable |
| 50.01 – 75.00 | Quite Sustainable |
| 75.01 – 100.00 | Very Sustainable |

The results of the analysis from the MDS will give the attribute of a sustainability booster. The attribute of the reference will be an advanced analysis factor in the assessment of the development planning of the corn farming area. The objective of this integration is to obtain weighting, scores, and values of criteria and sub-criteria that are balanced against each aspect in the assessment of alternative planning such as Ecological sustainability ($S_{Ecological}$), economic sustainability ($S_{Economic}$) and social sustainability ($S_{Social}$).

**Sustainable corn farming area development planning analysis**

Sustainable corn development planning analysis is carried out through a spatial statistical analysis approach by performing mathematical calculations of the sustainability of ecological, economic and social aspects with simple logarithms as follows:

S-$_{Ecological}$ = Sub criteria score x Class Value
S-$_{Economic}$ = Sub criteria score x Class Value
S-$_{Social}$    = Sub criteria score x Class Value

The result of the above formula is then done raster analysis of Sustainable Corn Farming Area Development Planning (SP-Corn) with the following equation:

SP-$_{Corn}$ = (S-$_{Ecological}$ weight x S-$_{Ecological}$) + (S-$_{Economic}$ weight x S-$_{Economic}$) + (S-$_{Social}$ weight x S-$_{Social}$)

The equation yields value from the combined integration of sustainability pillars, which will then be divided into three priority classes of development. The categorization is based on natural breaks, the most accurate method for tabular data and has a picture of the smallest error volume based on the natural interval patterns of the data [17,18]. This method calculates the Goodness of Variance Fit (GVF) with the following equation [19]:

$$GVF = \frac{SDAM - SDCM}{SDAM} = 1 - \frac{SDCM}{SDAM}$$

SDAM (Sum of squared Deviations for Array Mean) indicates variance between classes; SDCM (Sum of squared deviations for Class Means) declares variance for each class.

Priority planning of agricultural area development is structured using descriptive analysis by making several priority directions in line with the adequacy analysis of sustainable agricultural development planning.

## RESULTS AND DISCUSSION

### Land Suitability Analysis

Land suitability analysis is an assessment of land quality regarding the suitability of cultivated crops, which influences its potential use as opportunities and challenges in managing and optimizing land productivity. The research results indicate that land suitability class S3 (Marginally Suitable) dominates in the study area with two subclasses: S3rf (2,303.30 ha) and S3rfn (559.43 ha).

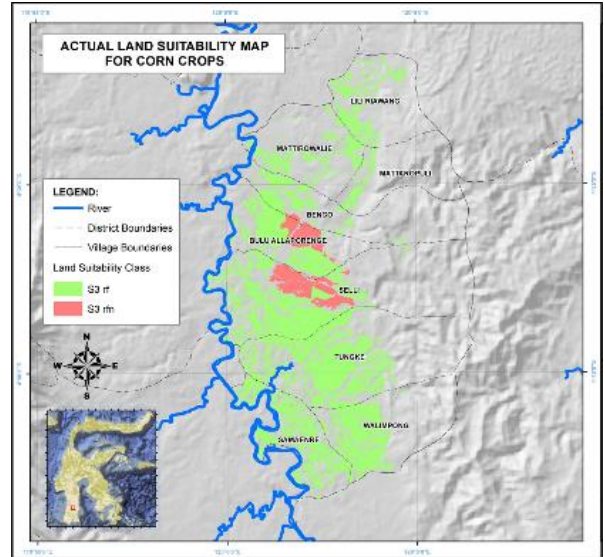

**Fig. 1.** Map of Actual Land Suitability for Corn Crops

This land suitability encompasses various combinations of limiting factors such as drainage (r), base saturation (f), and K20 (n). Impaired drainage can lead to harmful waterlogging for plant growth [20] and limit the choice of crops that can be grown. High base saturation can cause various soil nutrition and physical problems that hinder root development and nutrient absorption. Imbalanced K2O content can disrupt plant nutrient balance and affect productivity [21]. Therefore, careful management strategies are required to address these limiting factors through effective drainage management, soil pH adjustment, or appropriate fertilization to improve land conditions and enhance corn production potential [22, 23].

**Table 2.** Actual Land Suitability Class Values for Corn Cultivation

| Land quality/characteristics | SPL 1 | Class | SPL 2 | Class | SPL 3 | Class | SPL 4 | Class | SPL 5 | Class | SPL 6 | Class |
|---|---|---|---|---|---|---|---|---|---|---|---|---|
| **Temperature** | | | | | | | | | | | | |
| Annual average (C) | 25,8 | S1 | 25,8 | S1 | 25,8 | S1 | 25,8 | S1 | 25,8 | S1 | 25,8 | S1 |
| **Water availability (w)** | | | | | | | | | | | | |
| Dry Month (<75mm) | 3 | S1 | 3 | S1 | 3 | S1 | 3 | S1 | 3 | S1 | 3 | S1 |
| Annual rainfall (mm) | 1872,8 | S1 | 1872,8 | S1 | 1872,8 | S1 | 1872,8 | S1 | 1872,8 | S1 | 1872,8 | S1 |
| Humidity | 82,40% | S1 | 82,40% | S1 | 82,40% | S1 | 82,40% | S1 | 82,40% | S1 | 82,40% | S1 |
| **Rooting medium (r)** | | | | | | | | | | | | |
| Soil drainage | Hampered | S3 | Hampered | S3 | Hampered | S3 | Hampered | S3 | Hampered | S3 | Hampered | S3 |
| Soil texture | Liat | S1 | Liat | S1 | Liat | S1 | Liat | S1 | Liat | S1 | Liat | S1 |
| Effective depth (cm) | 50 cm | S1 | 50 cm | S1 | 50 cm | S1 | 50 cm | S1 | 50 cm | S1 | 50 cm | S1 |
| **Nutrient retention (f)** | | | | | | | | | | | | |
| KTK | 25,21 | S1 | 28,42 | S1 | 24,53 | S1 | 24,86 | S1 | 25,27 | S1 | 26,2 | S1 |
| Base Saturation (%) | 30 | S3 | 21 | S3 | 33 | S3 | 35 | S2 | 22 | S3 | 30 | S3 |
| Soil pH | 6,01 | S3 | 5,98 | S3 | 6,12 | S3 | 6,85 | S3 | 5,68 | S3 | 6,4 | S3 |
| C-Organic | 1,57 | S1 | 0,87 | S1 | 1,68 | S1 | 2,24 | S1 | 0,84 | S1 | 1,96 | S1 |
| **Nutrients available (n)** | | | | | | | | | | | | |
| N-Total (%) | 0,22 | S1 | 0,18 | S1 | 0,24 | S1 | 0,31 | S1 | 0,17 | S1 | 0,23 | S1 |
| $P_2O_5$ | 10,42 | S2 | 9,6 | S2 | 14,71 | S2 | 15,12 | S2 | 10,48 | S2 | 10,07 | S2 |
| $K_2O$ | 0,33 | S2 | 0,44 | S3 | 0,34 | S2 | 0,14 | S2 | 0,21 | S2 | 0,36 | S2 |
| **Erosion hazard level (e)** | | | | | | | | | | | | |
| Danger of erosion | Very Light | S1 | Very Light | S1 | Very Light | S1 | Very Light | S1 | Very Light | S1 | Very Light | S1 |
| Slope | < 8 % | S1 | < 8 % | S1 | < 8 % | S1 | < 8 % | S1 | < 8 % | S1 | < 8 % | S1 |
| **Flood danger (b)** | F0 | S1 | F0 | S1 | F0 | S1 | F0 | S1 | F0 | S1 | F0 | S1 |
| **Actual Land Suitability Class** | | S3 rf | | S3 rfn | | S3 rf | | S3 rf | | S3rf | | S3 rf |

## Multidimensional Scaling Analysis for Sustainable Corn Agricultural Development Planning

The data collection process through literature review and in-depth analysis has resulted in several subcriteria from the ecological, economic, and social aspects. These subcriteria are utilized in the sustainability analysis of corn cultivation using the *Rap-Corn* ordination technique. This method is a statistical technique involving multidimensional transformation [24]. The analysis can better illustrate the distribution and factors influencing the sustainability of corn cultivation in the study area. This ordination technique aids in understanding the complex patterns associated with sustainability aspects [25].

Based on the results of the *Rap-Corn* analysis on the radar chart diagram, the ecological dimension depicts less sustainability with an index value of 49.40% (Fig. 3a), indicating significant ecological imbalance and vulnerability. Meanwhile, the economic dimension index has a value of 61.69% (Fig. 3b), and the social dimension index is 61.26% (Fig. 3c), indicating relatively sustainable economic growth and improved community well-being. These results suggest that although some aspects achieve a relatively sustainable level, further efforts are needed to enhance ecological balance and reduce economic and social inequality to achieve a more holistic sustainability.

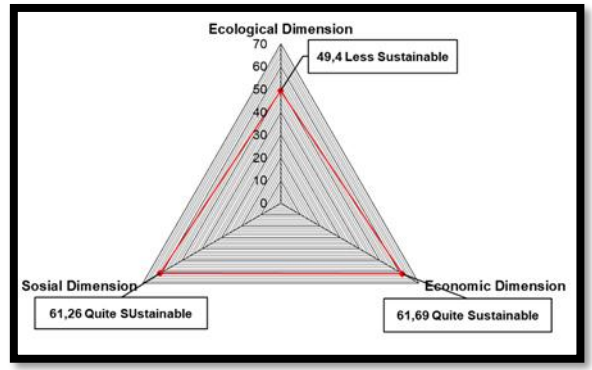

**Fig. 2.** Sustainability Kite Diagram for Corn Agricultural Area Planning

The assessment of the ecological dimension sustainability status involves 11 attributes (Fig. 3d). Leveraging attributes related to the Ecological Dimension Index shows a Root Mean Square (RMS) value of 3.46 for soil organic carbon content suitability, 3.06 for biological agent utilization, and 1.3 for fertilization, which play crucial roles in determining the ecological sustainability of an agricultural system. These attributes are sensitive factors that need to be considered for improvement in practices supporting environmental sustainability. Improvements in fertilization and the use of organic materials can help reduce soil degradation, enhance crop productivity, and achieve better and more sustainable ecological balance [22,23].

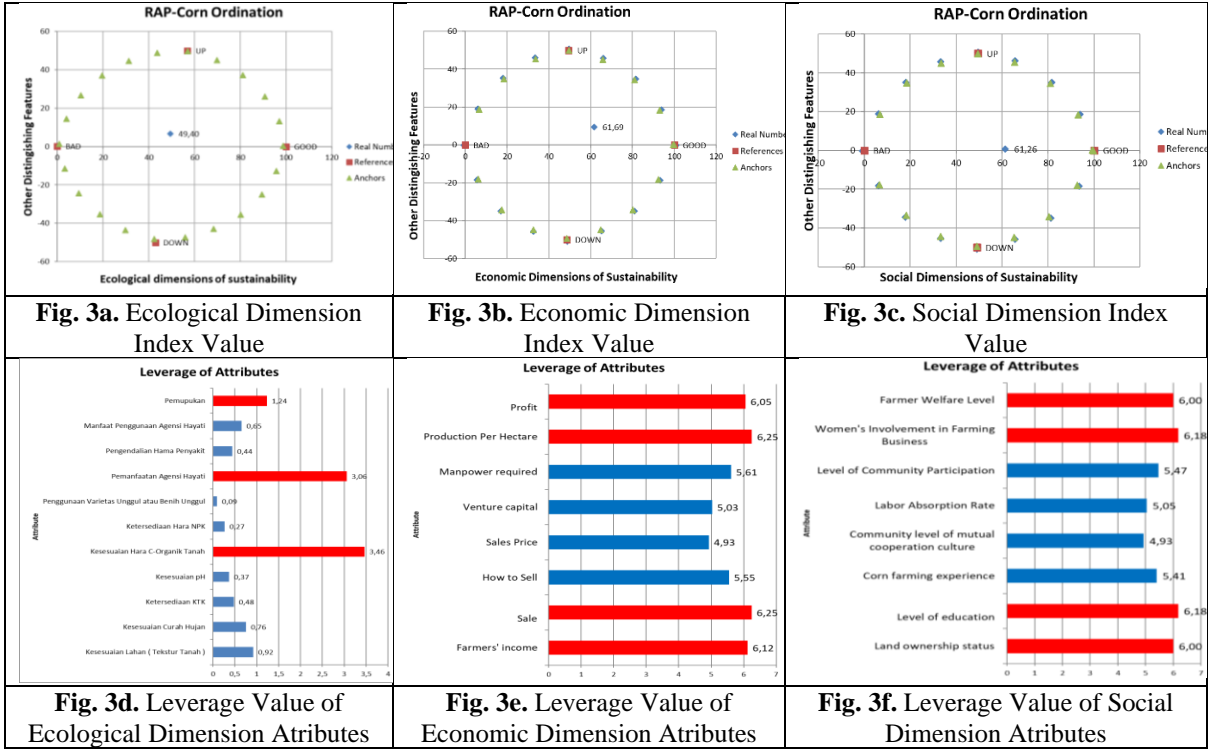

| **Fig. 3a.** Ecological Dimension Index Value | **Fig. 3b.** Economic Dimension Index Value | **Fig. 3c.** Social Dimension Index Value |
|---|---|---|
| **Fig. 3d.** Leverage Value of Ecological Dimension Atributes | **Fig. 3e.** Leverage Value of Economic Dimension Atributes | **Fig. 3f.** Leverage Value of Social Dimension Atributes |

The leveraged attributes related to the Economic Dimension (Fig. 3e) with an RMS value of 6.05 for profit, 6.25 for production per hectare, 6.25 for sales, and 6.12 for farmer income are crucial sensitive factors. Profitability is the main focus for farmers and increasing production per hectare is key to boosting agricultural product sales, thus increasing farmers' income [26]. Therefore, to enhance economic welfare in the agricultural sector, it is essential to consider and improve performance in terms of profit, production per hectare, sales, and farmer income.

The leveraged attributes from the social dimension (Fig. 3f) show an RMS value of 6.00 for farmer welfare level. Farmer welfare level emphasizes the need for farmers to have sufficient access to resources and services that support their well-being [27]. Women's involvement in agricultural business with a value of 6.18 highlights the crucial role of women in the agricultural sector and the overall economy [28]. The education level with a value of 6.18 emphasizes that higher education can provide farmers with broader opportunities for more sustainable and efficient farming practices [29]. Meanwhile, the RMS value for land ownership status is 6.0. Land ownership status can influence farmers' active involvement in managing and developing sustainable agricultural land [30]. Therefore, to improve social conditions and farmer welfare, it is essential to consider and enhance performance in terms of farmer welfare, women's involvement in agricultural business, education level, and land ownership status.

**Analysis of Weights and Scores of Ecological, Economic, and Social Aspect Criteria**

Integration of ecological, economic, and social aspects is an approach that considers the impact of activities or policies on the environment, economy, and society simultaneously. The Analytic Hierarchy Process (AHP) method is used to obtain scores and weights for relevant criteria in each attribute that acts as leveraged factors for sustainability from the results of the MDS analysis. The AHP results yield the weight values for each criterion (Fig. 4), with the ecological aspect having a weight of 0.46, the economic aspect with a weight of 0.31, and the social aspect obtaining a weight of 0.21.

The ecological aspect consists of 3 criteria: fertilization, organic material utilization, and soil organic carbon content suitability. The economic aspect includes 4 criteria: profit, production per hectare, sales per hectare, and farmer income per month. Meanwhile, the social aspect consists of 4 criteria, namely farmer welfare level, women's involvement in agricultural business, education level, and land ownership status. The score values for subcriteria are elaborated in the following figures and explanations.

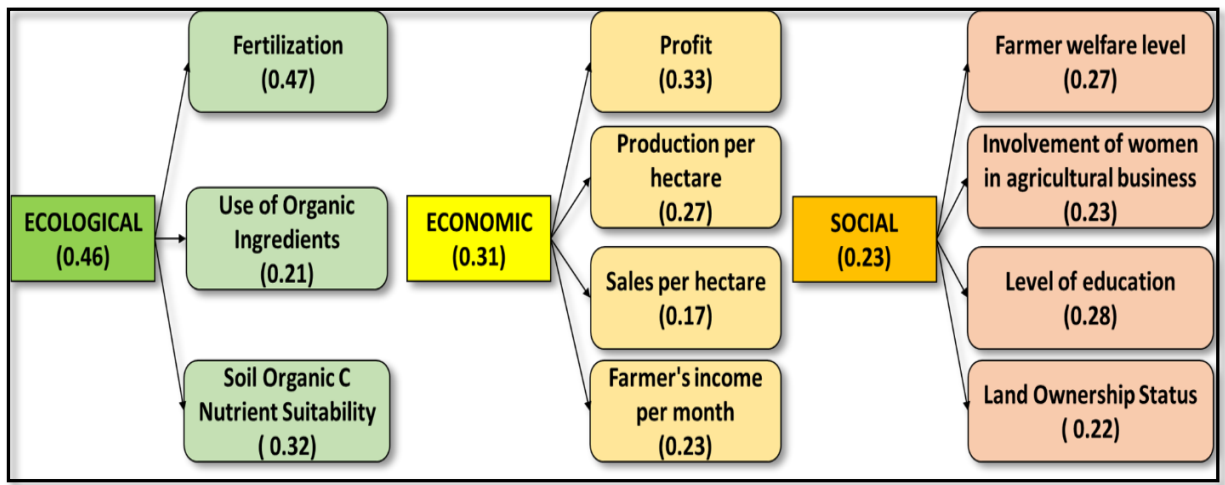

**Fig. 4.** Criteria Weights and Subcriteria Scores for Ecological, Economic, and Social Aspects

- **Ecological Aspect**

The AHP results yield scores for each indicator that correspond to their importance in achieving the sustainability goals of corn farming. Fertilization obtains a score of 0.47, reflecting the effectiveness of fertilizer use in corn farming. Proper fertilizer use can minimize negative environmental impacts [31,32]. The utilization of organic materials receives a score of 0.21, indicating the extent to which organic materials are utilized in corn fields. The use of organic materials can enhance soil fertility, reduce the need for chemical fertilizers, and overall improve soil quality [31]. Meanwhile, the suitability of soil organic carbon content receives a score of 0.32, evaluating the suitability of organic soil nutrients in supporting corn plant growth. Adequate nutrient availability in the soil is crucial for optimal plant growth and also plays a role in maintaining soil ecosystem balance [32]. The score values for the ecological subcriteria classes from the AHP results are presented in the following figure.

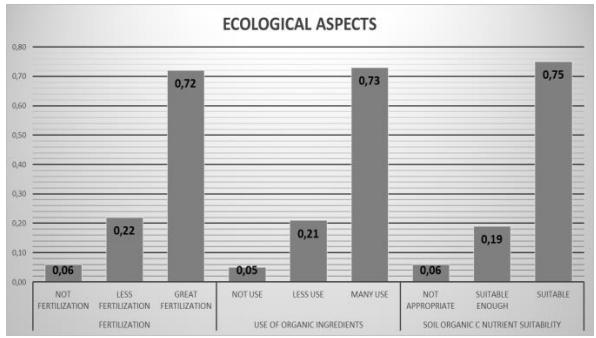

**Fig. 5.** The Score Values for Ecological Aspect Subcriteria Classes

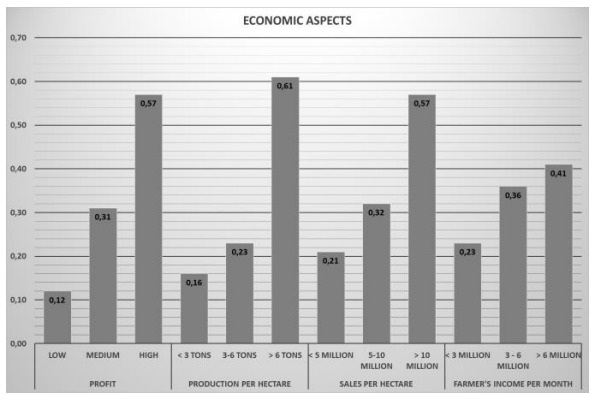

**Fig. 6.** The Score Values for Economic Aspect Subcriteria Classes

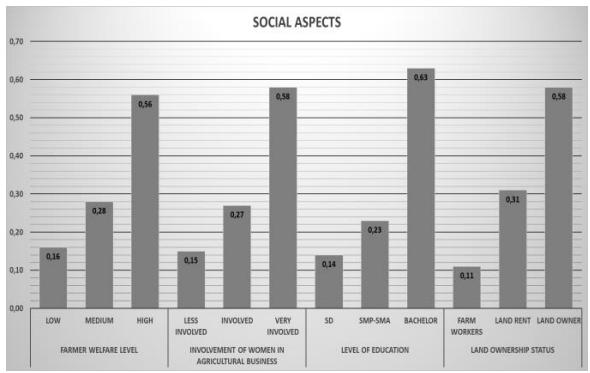

**Fig.7.** The Score Values for Social Aspect Subcriteria Classes

- **Economic Aspect**

The AHP results from the economic aspect provide insights into the score values of economic factors in the development of sustainable corn agriculture. The profit score obtained is 0.33, referring to the profitability of corn farming. High profits indicate that the business generates enough profit to support the sustainability of agricultural operations [33]. Production per hectare, with a score of 0.27, indicates the productivity of corn farming. High productivity is essential to ensure that the farming operation is efficient and capable of providing adequate yields to meet market demands [34]. The sales per hectare score is 0.17. A high sales volume indicates effective marketing of products [35]. Farmer's income per month, with a score of 0.23, evaluates whether the income obtained by farmers each month is sufficient for their welfare. The score values for the economic subcriteria classes from the AHP results are presented in the following figure.

- **Sosial Aspect**

The AHP results from the social dimension include four main indicators, each with a respective score. Farmer welfare obtains a score of 0.27, which refers to the importance of good farmer welfare levels in creating a productive work environment and supporting long-term sustainability in agriculture [36]. Involvement of women in agricultural business receives a score of 0.23, evaluating the extent to which women can enhance productivity and sustainability in agricultural operations. Level of education, with a score of 0.28, reflects how higher education levels can assist farmers in adopting more innovative and sustainable farming practices [29]. Meanwhile, land ownership status receives a score of 0.22, indicating that clear and secure ownership status is crucial for providing legal certainty to farmers and encouraging long-term investments in agricultural development [37]. The score values for the social subcriteria classes from the AHP results are presented in the following figure.

**Analysis of Sustainable Corn Agricultural Development Planning**

Sustainable corn agricultural development planning aims to create a balanced integration of ecological sustainability, economic profitability, and social welfare. Based on the AHP results, weights for criteria and scores for subcriteria in the ecological, economic, and social aspects have been obtained.

**Tabel 3.** Bobot Kriteria, Skor Sub Kriteria dan kelas Perencanaan Pengembangan Kawasan Pertanian Jagung Berkelanjutan

| Criteria | Weight | Sub Criteria | Score | Class | Score |
|---|---|---|---|---|---|
| Ecological Aspects | 0,46 | Fertilization | 0.47 | Not Fertilization | 0,06 |
|  |  |  |  | Less Fertilization | 0,22 |
|  |  |  |  | Great Fertilization | 0,72 |
|  |  | Use of Organic Ingredients | 0.21 | Not Use | 0,05 |
|  |  |  |  | Less Use | 0,21 |
|  |  |  |  | Many Use | 0,73 |
|  |  | Soil Organic C Nutrient Suitability | 0.32 | Not Appropriate | 0,06 |
|  |  |  |  | Suitable enough | 0,19 |
|  |  |  |  | Suitable | 0,75 |
| Economic Aspects | 0,31 | Profit | 0.33 | Low | 0,12 |
|  |  |  |  | medium | 0,31 |
|  |  |  |  | High | 0,57 |
|  |  | Production per hectare | 0.27 | < 3 tons | 0,16 |
|  |  |  |  | 3-6 tons | 0,23 |
|  |  |  |  | > 6 tons | 0,61 |
|  |  | Sales per hectare | 0.17 | < 5 million | 0,21 |
|  |  |  |  | 5-10 million | 0,32 |
|  |  |  |  | > 10 million | 0,57 |
|  |  | Farmer's income per month | 0.23 | < 3 million | 0,23 |
|  |  |  |  | 3 - 6 million | 0,36 |
|  |  |  |  | > 6 million | 0,41 |
| Sosial Aspects | 0,23 | Farmer welfare level | 0.27 | Low | 0,16 |
|  |  |  |  | medium | 0,28 |
|  |  |  |  | High | 0,56 |
|  |  | Involvement of women in agricultural business | 0.23 | Less Involved | 0,15 |
|  |  |  |  | Involved | 0,27 |
|  |  |  |  | Very Involved | 0,58 |
|  |  | Level of education | 0.28 | SD | 0,14 |
|  |  |  |  | SMP-SMA | 0,23 |
|  |  |  |  | Bachelor | 0,63 |
|  |  | Land Ownership Status | 0.22 | Farm Workers | 0,11 |
|  |  |  |  | Land Rent | 0,31 |
|  |  |  |  | Land owner | 0,58 |

With the weights for criteria and scores for subcriteria as shown in Table 3, the formulation equation for sustainable corn agricultural development planning can be written as follows.

S-$_{Ecological}$ = (0,47 x *Fertilization Class Value*) + (0,21 x *Use of Organic Ingredients Class Value*) + (0,32 x *Soil Organic C Nutrient Suitability Class Value*)

S-$_{Economic}$ = (0,33 x *Profit Class Value*) + (0,27 x *Production per hectare Class Value*) + (0,17 x *Sales per hectare Class Value*) + (0,23 x *Farmer's income per month Class Value*)

S-$_{Social}$ = (0,27 x *Farmer welfare level Class Value* + (0,23 x *Involvement of women in agricultural business Class Value*) + (0,28 x *Level of education Class Value*) + (0,22 x *Land Ownership Status Class Value*)

SP-$_{Corn}$ = (0,46 x S-$_{Ecological}$) + (0,32 x S-$_{Economic}$) + (0,21 x S-$_{Social}$)

The equation above yields cumulative values, which are spatially represented by the distribution of ecological sustainability values ranging from 0.41 to 0.55 (Fig. 8a), economic sustainability values ranging from 0.30 to 0.49 (Fig. 8b), and social sustainability values ranging from 0.32 to 0.40 (Fig. 8c). The cumulative equation results of the three pillars of sustainable development yield a distribution of values ranging from 0.36 to 0.49, which are categorized into 3 priority classes using the natural breaks method. The first priority class ranges from 0.44 to 0.49, the second priority class ranges from 0.39 to 0.44, and the third priority class ranges from 0.36 to 0.39. These are spatially presented in Figure 9.

The prioritization grouping is made as a strategic step towards achieving the goals by formulating development directions from the results of previous analytical stages in a structured, integrated, and sustainable manner. The Prioritized Directions for Sustainable Corn Agricultural Development Planning are presented in Table 4.

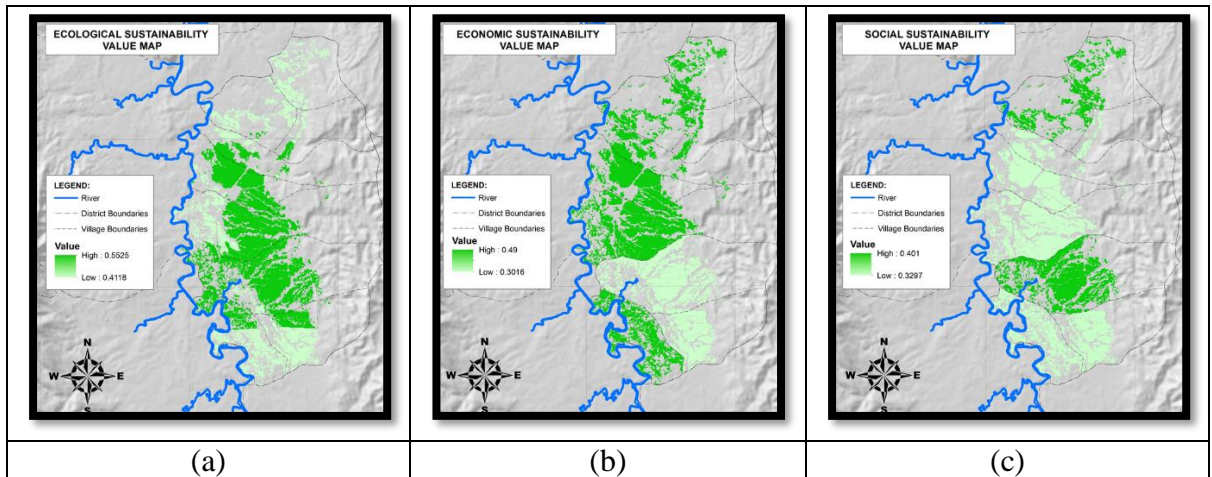

|(a)|(b)|(c)|

**Fiq. 8.** Map of the distribution of ecological, economic and social sustainability values

**Tabel 4.** Prioritized Directions for Sustainable Corn Agricultural Development Planning

| Priority | Development Directions |
|---|---|
| First Priority | • Increase soil organic matter levels with organic fertilizer<br>• The need for support and collaboration between government, research institutions and the private sector |
| Second Priority | • Increase soil organic matter levels with organic fertilizer<br>• The need for capacity development and empowerment programs<br>• The need for support and collaboration between government, research institutions and the private sector |
| Thierd Priority | • Increase soil organic matter levels with organic fertilizer<br>• The need for capacity development and empowerment programs<br>• The availability of educational programs<br>• The need for support and collaboration between government, research institutions and the private sector |

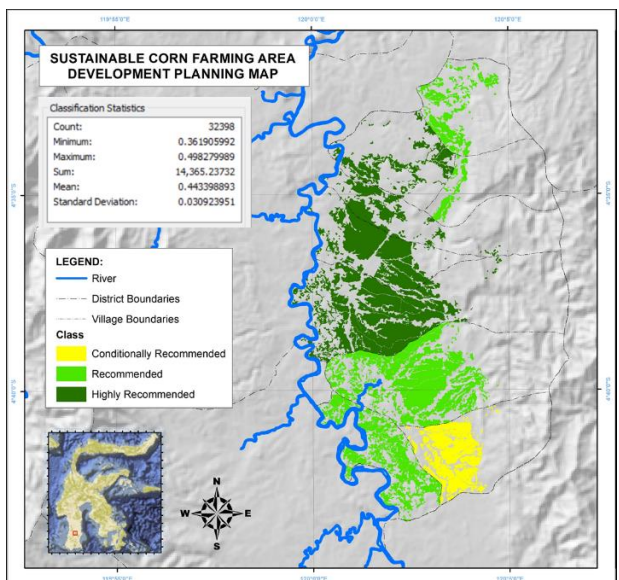

**Fig. 9.** Sustainable Corn Farming Area Development Planning Map

## CONCLUSION

Evaluating land suitability reveals that the S3 class (Marginally Suitable) predominates in the research area, with limiting factors including obstructed drainage, high base saturation levels, and imbalanced $K_2O$ content.

Sustainability evaluation through *RAP-Corn* analysis indicates that the ecological dimension is less sustainable, while the economic and social dimensions are relatively sustainable. Further efforts are required to enhance ecological balance and improve economic and social welfare to achieve sustainable development.

The AHP analysis results provide relevant weight and score distributions, depicting clear preferences. The multicriteria analysis model, integrating multidimensional scaling and the analytic Hierarchy Process, offers a holistic, structured and informed approach as an alternative for sustainable corn agricultural planning. The outcomes of this model can formulate prioritized directions for sustainable corn agricultural development planning.